\documentstyle[12pt,aasms4]{article}

\def\lesssim{\mathrel{\hbox{\rlap{\hbox{\lower4pt\hbox{$\sim$}}}\hbox{$<$}}}}
\def\gtrsim{\mathrel{\hbox{\rlap{\hbox{\lower4pt\hbox{$\sim$}}}\hbox{$>$}}}}

\begin{document}

\title{Highly Viscous Accretion Disks with Advection}
 \author{I.V.Artemova\altaffilmark{1},
 \affil{Theoretical Astrophysics Center, Juliane Maries Vej 30, DK-2100, Copenhagen
\O, Denmark}
 G.S. Bisnovatyi-Kogan\altaffilmark{2},
 \affil{Space Research Institute, Profsoyuznaya 84/32, 117810 Moscow, Russia}
 G. Bj\"{o}rnsson\altaffilmark{3},
 \affil{Science Institute, Dunhagi 3, University of Iceland, IS-107 Reykjavik,
Iceland}
 I.D. Novikov\altaffilmark{4} }
 \affil{Theoretical Astrophysics Center, Juliane Maries Vej 30, DK-2100, Copenhagen
\O, Denmark}
 \affil{University Observatory, Juliane Maries Vej 30, DK-2100, Copenhagen \O,
Denmark}
 \affil{NORDITA, Blegdamsvej 17, DK-2100 Copenhagen \O, Denmark}
 \affil{Astro Space Center of P.N. Lebedev Physical Institute, Profsoyuznaya
84/32,
 	117810 Moscow, Russia}
 \altaffiltext{1}{e-mail:julia@nordita.dk}
 \altaffiltext{2}{e-mail:Gkogan@mx.iki.rssi.ru}
 \altaffiltext{3}{e-mail:gulli@raunvis.hi.is}
 \altaffiltext{4}{e-mail:novikov@nordita.dk}

\begin{abstract}
We consider the effects of advection and radial gradients of pressure
and radial drift velocity on the structure of optically thick accretion
disks. We concentrate our efforts on highly viscous disks, $\alpha=1.0$,
with large accretion rates. Contrary to disk models neglecting advection,
we find that continuous solutions extending from the outer disk regions 
to the inner edge exist for all accretion rates we have considered. 
We show that the sonic point moves outward with increasing accretion rate,
and that in the innermost disk region advection acts as a heating process
that may even dominate over dissipative heating. Despite the importance of
advection on it's structure, the disk remains geometrically thin. 
\end{abstract}
 

\section{Introduction}
\noindent
The ``standard accretion disk model'' of Shakura (1972) and Shakura \& 
Sunyaev (1973), that has been widely used to model accretion flows around 
black holes, is based on a number of simplifying assumptions. In particular, 
the flow is assumed to be geometrically thin and with a Keplerian 
angular velocity distribution. This assumption allows gradient terms in the 
differential equations describing the flow to be neglected, reducing them 
to a set of algebraic equations, and thereby fixes the angular momentum 
distribution of the flow. For low accretion rates, $\dot M$, this assumption 
is generally considered to be reasonable.

Since the end of the seventies, however, it has been realized that for high 
accretion rates, advection of energy with the flow can crucially modify 
the properties of the innermost parts of accretion disks around black holes. 
A deviation from a Keplerian rotation may result.

Initial attempts to solve the more general disk problem only included advection
of energy and the radial gradient of pressure in models with small values of
the viscosity parameter, $\alpha=10^{-3}$ (Paczy\'nski \& Bisnovatyi-Kogan 1981),
and it was shown that including radial velocity in the radial momentum equation 
would not change principally the results for such a small $\alpha$ 
(Muchotrzeb \& Paczy\'nski 1982). Liang \& Thomson (1980) emphasised the importance 
of the transonic nature of the radial drift velocity, and
the influence of viscosity on the transonic accretion disk solutions was noted by 
Muchotrzeb (1983), who claimed that such solutions only existed for viscosity 
parameters smaller than $\alpha_{*}\simeq 0.02-0.05$. Matsumoto et al.\ (1984),
then showed that solutions with $\alpha>\alpha_{*}$ do in fact exist, but the 
nature of the singular point, where the radial velocity equals the sound velocity, 
is changing from a saddle to a nodal type and the position of this point is shifted 
substantially outwards in the disk. Matsumoto et al.\ (1984) also demonstrated the 
non-uniqueness of the solutions with a nodal type critical point for given 
Keplerian boundary conditions at the outer boundary of the disk (see also 
Muchotrzeb-Czerny 1986). Extensive investigation of accretion disk models 
with advection for a wide range of the disk parameters, $\dot{M}$ and $\alpha$, 
was conducted by Abramowicz et al.\ (1988), with special emphasis on low $\alpha$.
Misra \& Melia (1996) considered optically thin two-temperature disk models
and treated advection in the framework of the Keplerian disk model,
but fixed the proton temperature somewhat arbitrarily at the outer boundary.
Chakrabarti (1996) solved the advection problem containing shock waves near 
the innermost disk region, considering accretion through saddle points. 
Numerical solutions of accretion disks with advection have been obtained by 
Chen and Taam (1993) for the optically thick case with $\alpha=0.1$, and by 
Chen et al.\ (1996), for the optically thin case (see also Narayan 1996). 
A simplified account of advection has recently been attempted, either 
treating it like an additional algebraic term assuming a constant radial 
gradient of entropy (Abramowicz et al.\ 1995; Chen et al.\ 1995; Chen 1995), 
or using the condition of self-similarity (Narayan \& Yi 1994). 

Over the last few years it has become clear, that neglecting the 
advective heat transport at high $\dot{M}$ leads to qualitatively wrong 
conclusions about the topology of the family of solutions of the disk 
structure equations (see for example Abramowicz et al.\ 1995; 
Chen et al.\ 1995; Artemova et al.\ 1996).
The disk structure equations without advection give rise two branches of 
solutions: optically thick and optically thin, which do not intersect if 
$\dot M<\dot M_{b} \approx (0.6-0.9)  \dot{M}_{\rm Edd}$ for $\alpha=1$ 
and $M_{BH}=10^{8}M_{\odot}$, where $\dot{M}_{\rm Edd}$ is the Eddington 
accretion rate (Artemova et al.\ 1996). For larger accretion rates there 
are no solutions of these equations extending continuously from large to 
small radii, and with Keplerian boundary conditions at the outer boundary of 
the disk (see also Liang \& Wandel 1991; Wandel \& Liang 1991; 
Luo \& Liang 1994). It was argued by Artemova et al.\ (1996), that for 
accretion rates larger than $\dot{M}_b$, advection becomes critically 
important and would allow solutions extending all 
the way to the inner disk edge also to exist for $\dot{M}>\dot{M}_{b}$.

The goal of the present paper is to construct explicitly accretion disk 
models for high $\dot{M}$ and large $\alpha$ taking advective heat transport 
self-consistently into account. We also include radial gradients of pressure 
and radial drift velocity and we allow for the non-Keplerian character of the 
circular velocity. Furthermore, we use the geometrically thin disk approximation 
because, as will be seen in our solutions, the relative thickness of the disk 
is substantially less than unity. We show that solutions extending from large 
radii to the inner edge of the disk can be constructed even for 
accretion rates considerably larger than $\dot M_b$. We find that 
advection is very important in the innermost disk region, although the flow 
does not deviate strongly from Keplerian down to the region where the radial
inflow velocity approaches the local sound speed.

In \S 2 we introduce our model and describe our solution methods, while
in \S 3 we discuss our results.

\section{The Model and the Method of Solution}

\noindent
In this paper we will only consider optically thick solutions to the
disk equations. When advective cooling is important we assume that it
can be sufficiently well modelled by adding it self-consistently to 
other cooling mechanisms in a geometrically thin disk.

We use from now on geometric units with $G=1$, $c=1$, use $r$ as the radial 
coordinate scaled to $r_g=M$, and scale all velocities to $c$. 
We work with the pseudo-Newtonian potential proposed by Paczy\'nski 
and Wiita (1980), $\Phi=-M/(r-2)$, that provides an accurate, yet simple
approximation to the Schwarzschild geometry. 
We normalise the accretion rate as $\dot{m}=\dot M/\dot M_{\rm Edd}$, where
$\dot M_{\rm Edd}=L_{\rm Edd}=4\pi M m_p/\sigma_T$, in our units.

We use the same equations and ingredients in our models as in 
Artemova et al.\ (1996), except for changes required by the Paczy\'nski-Wiita
potential and the differential terms in the energy equation and the radial
momentum equation.
The following equations are therefore modified:

1. Conservation of angular momentum for a steady-state accretion in the $\alpha$-disk 
model, is written as
\begin{equation} 
\dot{m} \left(r_g\Omega\right) \frac{3}{2}\left|\frac{d\ln\Omega}{d\ln r}\right|^{-1}f 
= \left(\frac{\sigma_{\rm T}}{m_{\rm p}}\right)h \alpha P,
\label{eq:angmom}
\end{equation}
where, $\Omega$, is the angular velocity, the factor $f=1-l_{\rm in}/l$, where,
$l=r^{2}\Omega$, is the specific angular momentum and $l_{\rm in}$ is the value 
of $l$ lost from the disk at the innermost edge and swallowed by the black hole. 
The half thickness of the disk 
is denoted by $h$, and $P$ is the total pressure in the equatorial plane of the disk.

2. The energy equation has the form 
\begin{equation}
Q_{+}=Q_{\rm adv}+Q_{\rm loc},
\label{eq:energy}
\end{equation}
where $Q_{\rm loc}$ is in general the total rate of all local cooling processes 
(see Artemova et al.\ 1996),
and the viscous heating rate per unit area, $Q_{+}$, is given by the formula (see e.g. 
Bisnovatyi-Kogan 1989; Frank, King \& Raine, 1992)
\begin{equation}
\left(\frac{r_g\sigma_{\rm T}}{3 m_{\rm p}}\right)Q_{+}=\dot m\left(r_g\Omega\right)^2 
 \frac{2}{3}\left|\frac{d\ln\Omega}{d\ln r}\right| f.
\label{eq:qplus}
\end{equation}
The advective cooling rate can be written in the form (see e.g. Chen \& Taam 1993):
\begin{equation}
Q_{\rm adv}=-\frac{\dot M}{2\pi r} T \frac{dS}{dr}
=-\frac{\dot M}{2 \pi r}\left[\frac{dE}{dr} + P\frac{dv}{dr}\right],
\label{eq:qadv}
\end{equation}
where $T$ is the temperature and $S$ is the specific entropy. Here, $E$ is the energy 
per unit mass of the gas and $v=1/\rho$, where $\rho$ is the matter density.
With $Q_{\rm adv}$ of the form given in equation (\ref{eq:qadv}), the energy
balance becomes a differential equation. 

3. The momentum equation in the radial direction takes into account pressure and radial 
velocity gradients and is written in the form:
\begin{equation}
\frac{1}{\rho}\frac{dP}{dr}=(\Omega^{2}-\Omega_{\rm K}^{2})r-v_{r}\frac{dv_{r}}{dr},
\label{eq:radmom}
\end{equation}
where, $\Omega_{\rm K}=\sqrt{(\partial\Phi/\partial r)/r}$, is the Keplerian 
angular velocity in the Paczy\'nski-Wiita potential and $v_r$ is the radial 
drift velocity. Neglecting the gradient terms in equation (\ref{eq:radmom}), 
as is done in the standard model, fixes $\Omega=\Omega_{\rm K}$.

From equation (\ref{eq:radmom}) and mass conservation one gets:
\begin{equation}
\frac{d\ln v_{r}}{d\ln r}=\frac{a_{s}^{2}(1+(d\ln h)/(d\ln r))+
	r^{2}(\Omega^{2}-\Omega_{\rm K}^{2})}{v_{r}^{2}-a_{s}^{2}},
\label{eq:lnvr}
\end{equation}
where $a_{s}$ is defined as (Muchotrzeb \& Paczy\'nski 1981):
\begin{equation}
a_{s}^{2}=\left(\frac{d P}{d r}\right)\left(\frac{d \rho}{d r}\right)^{-1}
\end{equation}
Note, that $a_{s}$ is {\em not} a physical sound velocity but rather a formal quantity.
The vanishing of both the numerator and the denominator in equation (\ref{eq:lnvr}) 
at the same value of $r$, the "sonic point", provides the regularity condition 
required for a "transonic solution" of the flow structure. 

To solve the differential equations (\ref{eq:energy}) and (\ref{eq:radmom})
we adopt the following boundary conditions: At large radii, $r \gg 100$, 
the solution must coincide with the standard Keplerian disk solution obtained 
neglecting advection. In addition, the parameter $l_{\rm in}$ in 
equations (\ref{eq:angmom}) and (\ref{eq:qplus}), is an eigenvalue of 
the problem which is 
adjusted in such a way that the solution satisfies the regularity condition 
at the ``sonic point''. 

We solved this system of equations numerically by the method of subsequent 
iterations with fixed $\dot{m}$ and $\alpha$. Starting from the "standard 
disk" solution as the initial trial solution with a specific value of $l_{\rm in}$
in the function, $f=1-l_{\rm in}/l$, we then varied $l_{\rm in}$ to obtain 
a self-consistent solution. Typically the method converges to a solution 
after three to four iterations. 

In practice, we varied $l_{\rm in}$ in some interval and determined the positions 
of the points $r_{N}$ and $r_{D}$ where the numerator and the denominator 
of equation (\ref{eq:lnvr}) vanish, respectively. We then considered the 
dependence of 
the difference $(r_{D}-r_{N})$ on $l_{\rm in}$ and determined $l_{\rm in}$ for which 
the difference $(r_{D}-r_{N})$ is equal to zero. The corresponding $l_{\rm in}$ 
is an eigenvalue of the problem.  Examples of the dependences $(r_{D}-r_{N})$ 
on $l_{\rm in}$ are given in Figure 1 for $\dot{m}=10.0$ (accretion rate less 
than $\dot{m}_{b}$), and $\dot{m}=28.0$ (accretion rate substantially greater 
than $\dot{m}_{b}$).

For $\dot m>\dot m_b$, our method is very sensitive to the choice of the 
initial trial solution in the innermost disk region. Using a ``standard
disk'' solution down to $r=6$ is not possible, as for these large values of 
$\dot m$ there is no solution around $r\approx 13$, and the method cannot
bridge that gap to find a ``transonic'' solution. For our initial trial
solution, we therefore chose a value of $l_{\rm in}$ that allowed us to 
generate the trial solution down to small radii, and then iterated as 
described above. 

As is seen in Fig. 1, there are three values of $l_{\rm in}$ for each $\dot{m}$,
where $(r_{D}-r_{N})=0$, and some range of $l_{\rm in}$ where $(r_{D}-r_{N})$ 
is very close to zero.  Most likely that range corresponds to the nodal type of 
a sonic point at large $\alpha$, as obtained by Matsumoto et al.\ (1984) and others 
under some simplifications.  In this case the condition of regularity at the sonic 
point does not specify $l_{\rm in}$ uniquely. 
From our numerical method we are unable to determine if any $l_{\rm in}$ in 
the range of $l_{\rm in}$'s, where $r_D-r_N$ is close to zero, provides an
acceptable solution. 
Complete analysis of the character of the critical points needs a different
approach and will be performed elsewhere (but see next section).

Our method allows us to construct a self-consistent solution to the system of 
equations from very large radii, $r>100$, and down to the innermost regions of 
the disk.  The solution passes through a "sonic point" and continues closer 
towards the black hole.  But, we cannot construct the parts of the solutions 
in the very vicinity of the black hole where the angular velocity is very far 
from Keplerian and $l$ is almost constant. 
However, in all cases do we extend the solutions down to the value of $r$ 
at which $v_{r}$ becomes equal to the local adiabatic sound velocity. 
These radii are in general closer to the black hole than the location 
of the ``sonic point''.

\section{Results and Discussion}

\noindent
In Table 1 we summarise the parameters of the models for which $r_{D}-r_{N}=0$,
according to our computations.  For each fixed $\dot{m}$, the properties of the 
two (or three) self-consistent solutions are similar and differ only quantitatively.  
In all cases discussed below do we take $M_{BH}=10^{8}M_{\odot}$ and
$\alpha=1$.

We will now compare the solutions with and without advection.
In the ``standard model'', for accretion rates $\dot{m}<\dot{m}_{b}=14.315$, there 
always exist solutions that extend continuously from large to small radii. When 
$\dot{m}>\dot{m}_{b}=14.315$ there are no solutions in a range of radii around 
$r \approx 13$, and therefore no continuous solutions extending from large radii
to the innermost disk edge (see detailed discussion by Artemova et al.\ 1996,
where however, the Newtonian potential was used, resulting in $\dot{m}_{b}=9.4$).

In Figure 2a we plot the disk surface density $\Sigma=2\rho h$,
in an optically thick disk as a function of radius, $r$, in a model with
$\dot m=10$. The lowest curve is the solution of the standard model, the
upper ones are solutions number 2 and 3 in Table 1. Note that the solutions
including advection all terminate at radii considerably greater than $r=6$
(inner edge of the disk in the standard model).

In Fig.\ 2b we plot similarly the solutions for $\dot m=15$. In the standard
model, no solution exists in the region around $r\approx13$, but 
when advection is included, the structure 
of the solutions is completely different. Models 6 and 7 in Table 1 are shown.

For $\dot{m}<13$, including the gradient terms gives rather small corrections
to the standard disk model, see Figure 2a. When $\dot{m}>13$ advection becomes 
essential and for $\dot{m}>\dot{m}_{b}$ it changes the picture qualitatively. When 
$\dot{m}>\dot{m}_{b}$ solutions do exist extending continuously from 
large radii to the innermost disk region where the solution passes through
a "sonic point" (compare Figs 2a and 2b, see also Fig 4b below). 
As mentioned above (see the end of Section 2), we can extend our models only 
down to the region 
where the radial velocity becomes equal to the local adiabatic sound velocity. 
We are unable to calculate the properties of the flow for smaller radii.
Only more detailed analysis of this region (using other methods) allows one 
to determine the smoothness of the flow down to the event horizon of a black hole 
or verify the presence or absence of shocks in the region. 

In Figure 3 we plot a family of optically thick solutions for different $\dot{m}$,
clearly demonstrating that the solutions to the complete system of disk structure 
equations including advection and radial gradients have quite different properties 
at high $\dot{m}$ compared to the solutions of the standard disk model.

In Figure 4a we plot $Q_{\rm adv}/Q_+$ as a function of radius for $\dot m=10$ 
and $\dot m=28$, that bracket the cases we have studied. Outside the radius
where the entropy gradient is zero (and therefore $Q_{\rm adv}=0$, recall 
eq.\ [\ref{eq:qadv}]), advection provides an additional cooling, that is however, 
never substantial in our models. On the other hand, inside that radius, advection 
acts as a heating process that easily dominates over the dissipation rate that 
decreases rapidly near the inner edge of the disk (as $f\rightarrow 0$,
see eq.\ [\ref{eq:qplus}]).
Panels 4b and 4c show the 
corresponding Mach numbers and $h/r$-ratio, respectively. Note that although
the flow becomes transonic in the inner region, the disk can still be
considered geometrically thin.

In our calculations, the non-uniqueness of solutions at large $\alpha>\alpha_{*}$, 
passing through the critical point (Matsumoto et al.\ 1984; Muchotrzeb-Czerny 1986), 
is preserved.
It is sill not clear, if this non-uniqueness is a realistic physical fact 
which explanation may be highly problematic (see for example Kato et al.\ 1988,
where the authors argue that the fact that the transonic 
point is a nodal type critical point is equivalent to an instability condition), 
or is a result of restrictive precision of our numerical solutions.
Two possible approaches to clarify the situation can be suggested. In the 
first one, we could obtain an asymptotic solution of the disk equations near the 
gravitational 
radius and try to match it with the numerical solution going from the nodal point 
towards the inside. The second approach could be finding stationary solution by 
solving equations of non-stationary accretion with the appropriate boundary conditions.
Both approaches need substantial numerical work, that we plan to undertake in
the future. 

\acknowledgements

We would like to thank M.\ Abramowicz for very constructive and friendly criticism 
of the early version of the paper and helpful suggestions.  We gratitude A. Dolgov 
and A. Doroshkevich for discussion and E.\ Kotok for help. This paper was supported 
in part by the Danish Natural Science Research Council through grant 11-9640-1, 
in part by Danmarks Grundforskningsfond through its support for the establishment 
of the Theoretical Astrophysical Center. G.B.\ thanks NORDITA and TAC for partial 
support and hospitality, and acknowledges partial support form the Research Fund of
the University of Iceland.\ G.B.-K.\ thanks TAC and Astronomy Units of Queen Mary 
and Westfield College for partial support and hospitality during the initial 
stages of this work. This work was partially supported by the Nordic Project 
on Non-linear Phenomena in Accretion Disks around Black Holes.


\pagebreak
\begin{center}
FIGURE CAPTIONS
\end{center}
Fig. 1. Difference $(r_D - r_N)$ as a function of $l_{\rm in }$, the angular
momentum swallowed by the black hole. The 3 zero-points correspond to the
solutions satisfying the regularity condition (eq. [\ref{eq:lnvr}]), and the
corresponding $l_{\rm in}$ are the eigenvalues of the problem. (a) Accretion
rate of $\dot m=10$ and (b) $\dot m=28$.\\

Fig. 2. Disk surface density, $\Sigma=2\rho h$, as a function of radius, 
comparing solutions with and without advection. (a) Here, $\dot m=10$. 
The solid curve is the standard solution without advection. The 
dotted curve has $l_{\rm in}=3.782$ and the
dashed curve has $l_{\rm in}=4.025$
(models 2 and 3 in Table 1).
(b) The case $\dot m=15$. Again the solid curve is the standard model 
without advection. Notice the 'no solution' region around $r\approx 13$.
The dotted curve has $l_{\rm in}=4.125$ and the
dashed curve has $l_{\rm in}=4.513$ 
(models 6 and 7 in Table 1).\\

Fig. 3. Surface density, $\Sigma$, as a function of radius for $\alpha=1.0$ 
and different accretion rates. The solid curve is for $\dot m = 1.0$,
the long dashed, short dashed, dotted and dash-dotted curves have $\dot m= 
10, 15, 19$ and 28, respectively (models 2, 6, 10 and 13, respectively).

Fig. 4. Comparing the solutions for $\dot m=10$ (dashed curve) and 
$\dot m=28$ (solid curve) with $\alpha=1.0$, that bracket most of the 
cases we have studied. (a) Ratio of advective rate (eq. [\ref{eq:qadv}])
to viscous heating rate (eq. [\ref{eq:qplus}]). The advective rate equals
zero when the entropy gradient is zero (at $r\approx 18$ and $r\approx 40$,
for $\dot m=10$ and $\dot m=28$, respectively). Outside those radii, advection
provides rather small additional cooling in both cases. Inside these radii
advection acts as a strong source of heating. (b) Mach number and (c) ratio
$h/r$ for the same cases as in panel (a). The styles of the curves are the 
same as in panel (a).

\clearpage

\begin{deluxetable}{lrrr}
\tablewidth{27pc}
\tablecaption{Models  \label{tbl-1}}
\tablehead{
\colhead{N}               & \colhead{$\dot m$} &
\colhead{$l_{\rm in}$} & \colhead{$r_s$}    }

\startdata
1 & 10 & $ 3.570 $ & $9.90$ \nl
2 & 10 & $ 3.782 $ & $12.02$ \nl
3 & 10 & $ 4.025 $ & $13.95$ \nl
4 & 13 & $ 4.010 $ & $15.0$ \nl
5 & 13 & $ 4.170 $ & $16.4$ \nl
6 & 15 & $ 4.125 $ & $16.6$ \nl
7 & 15 & $ 4.513 $ & $19.9$ \nl
8 & 17 & $ 4.251 $ & $18.8$ \nl
9 & 17 & $ 4.850 $ & $23.9$ \nl
10 & 19 & $ 4.416 $ & $18.6$ \nl
11 & 19 & $ 5.088 $ & $25.8$ \nl
12 & 28 & $ 4.305 $ & $22.1$ \nl
13 & 28 & $ 5.169 $ & $30.6$ \nl
14 & 28 & $ 5.409 $ & $33.0$ \nl
\enddata
\end{deluxetable}

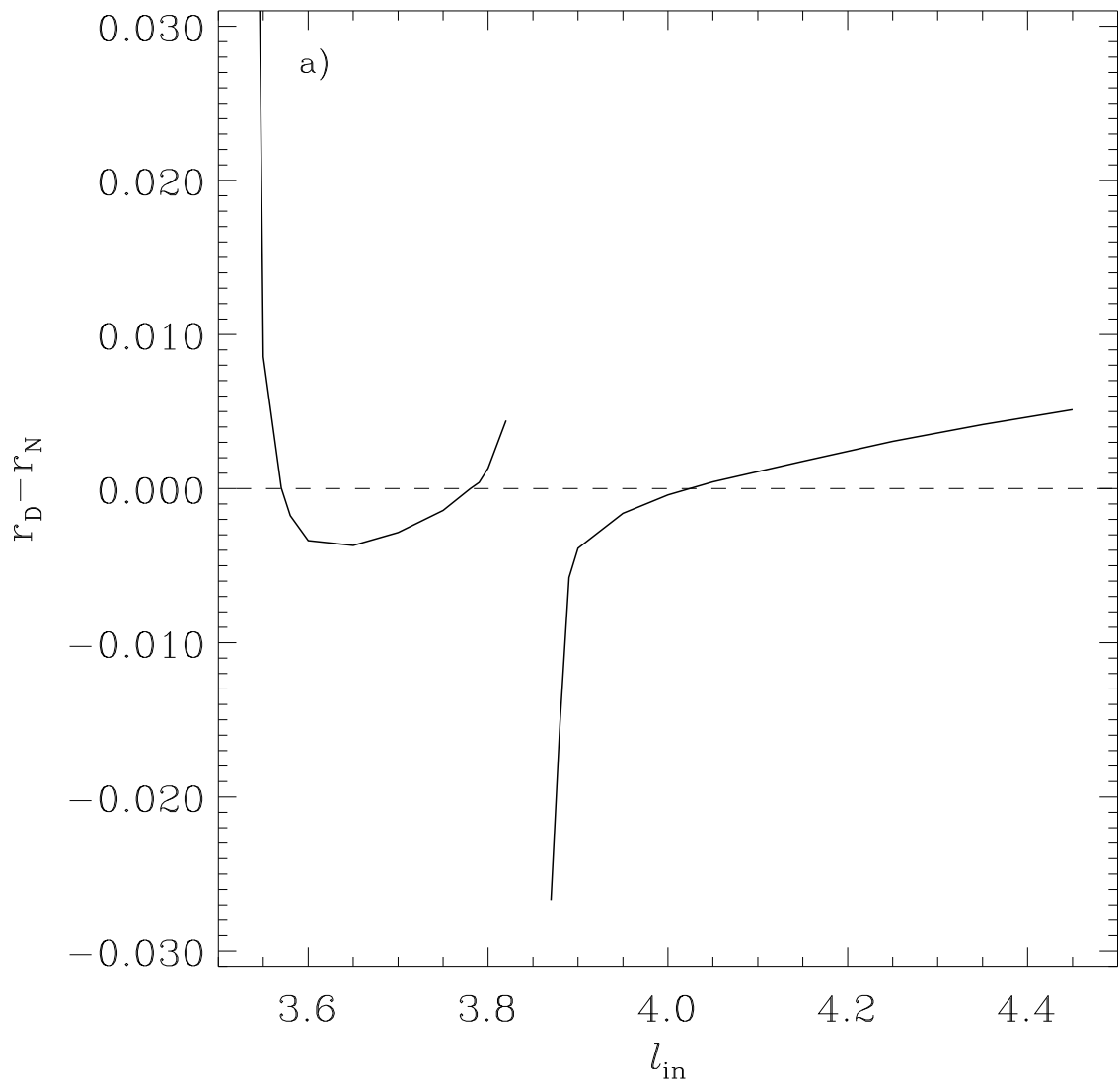
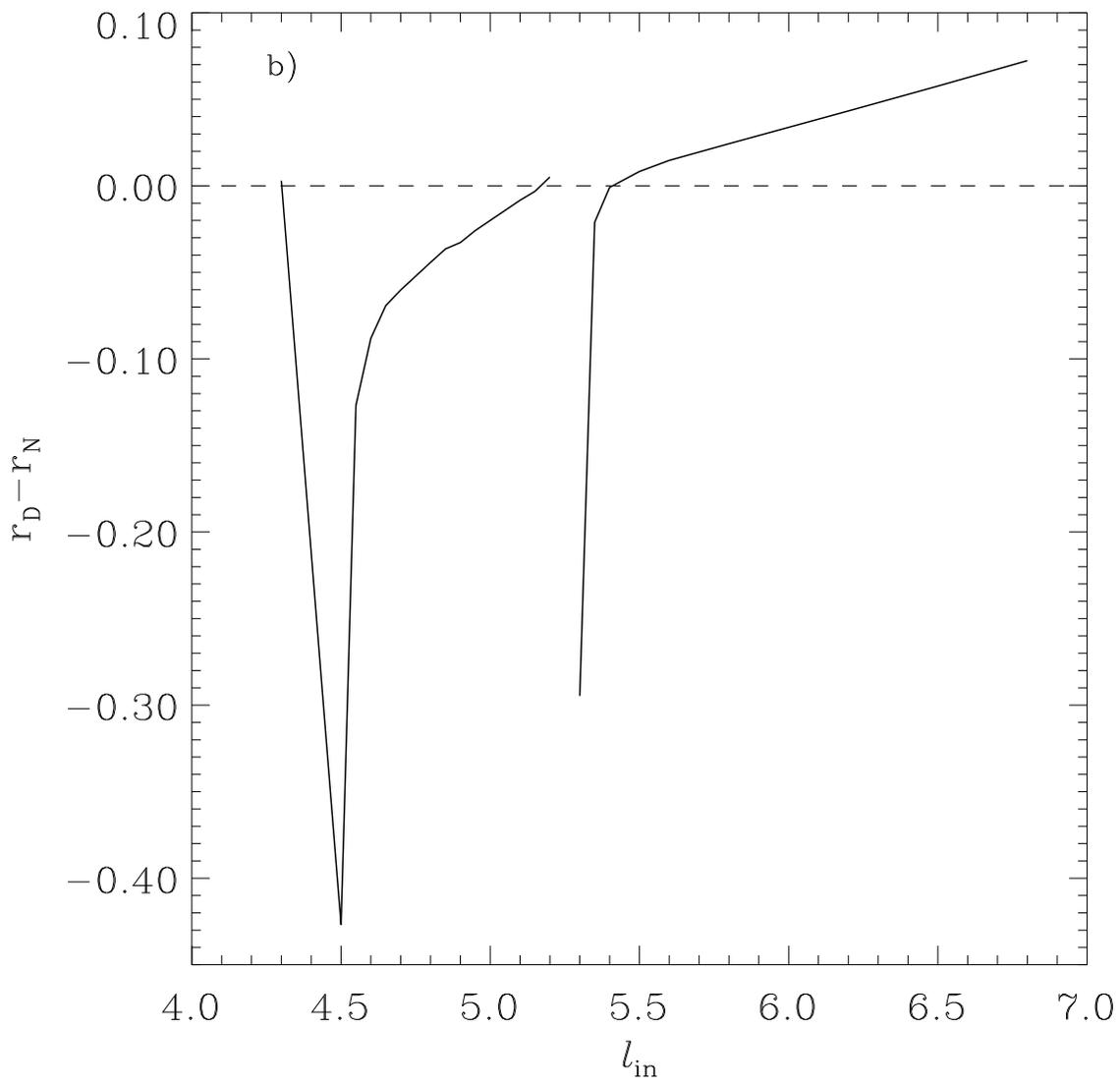
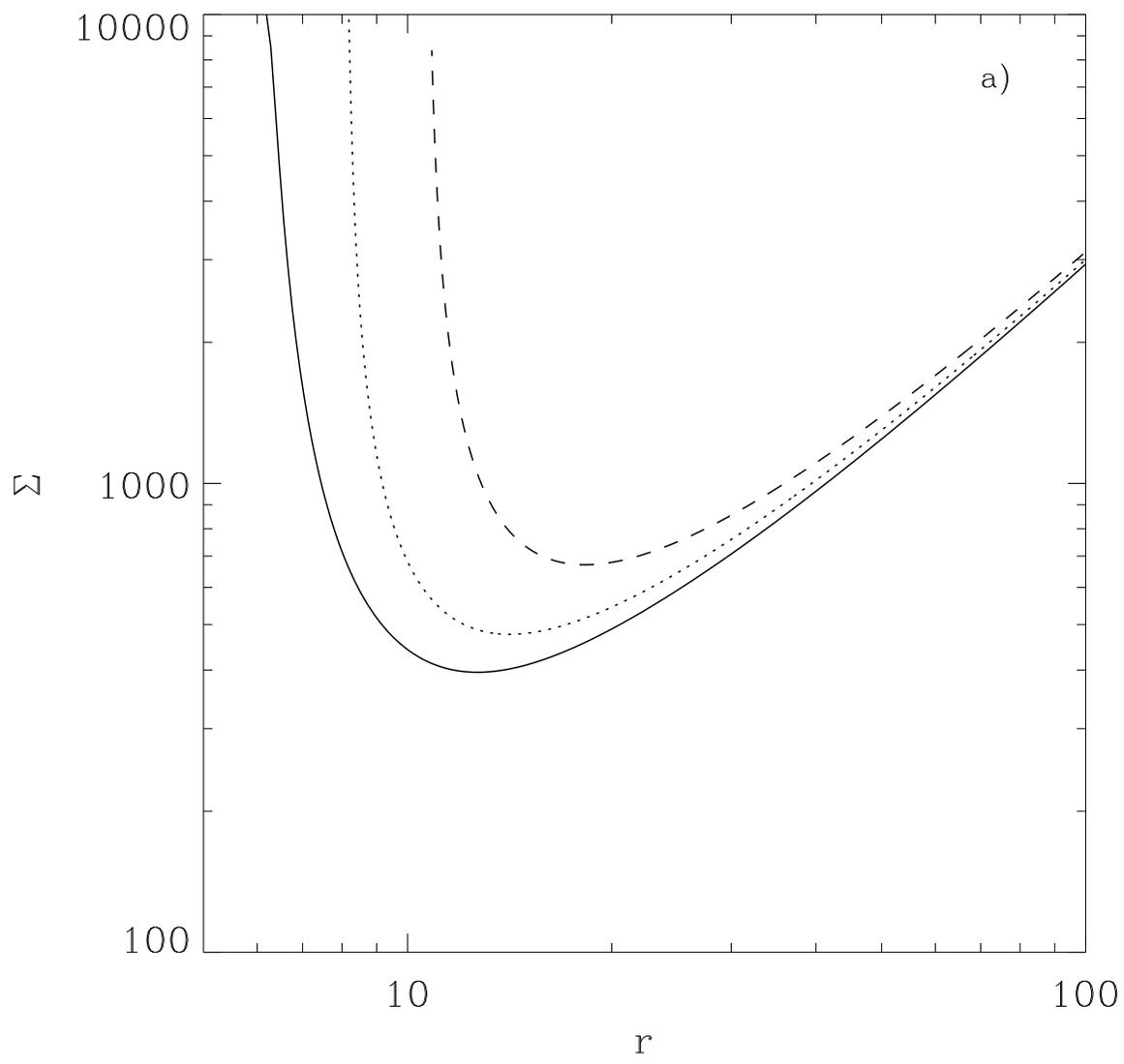
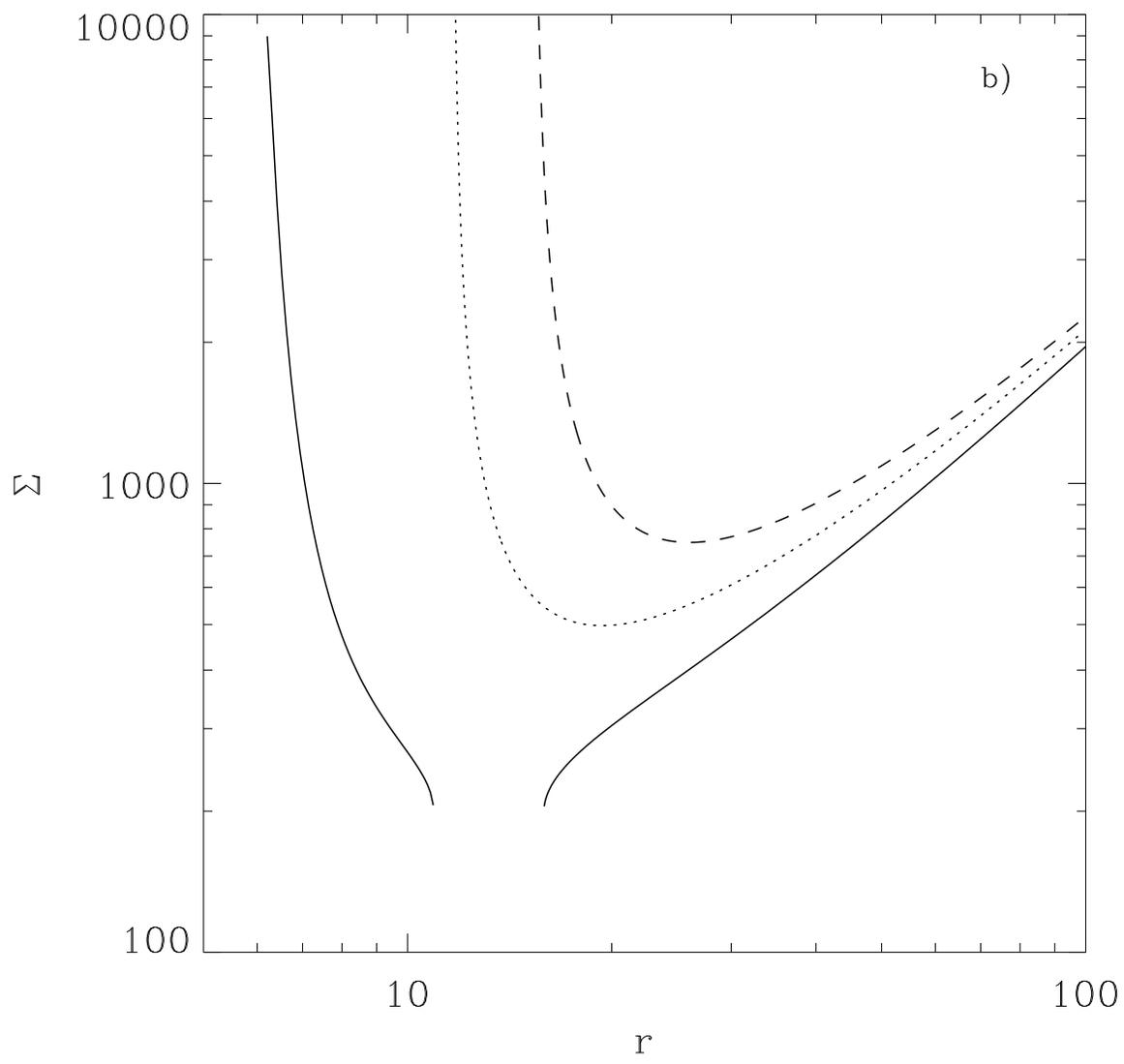
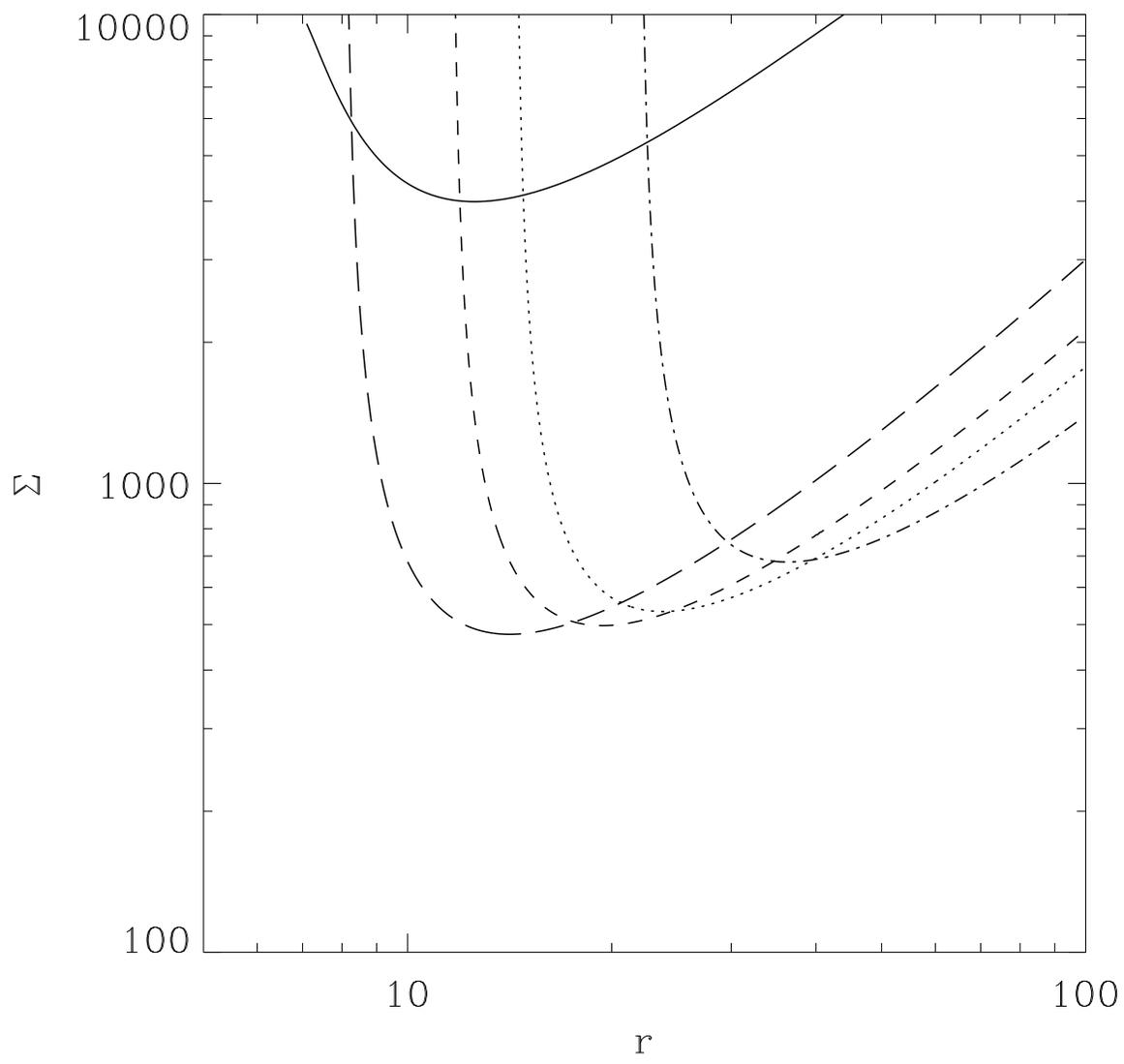
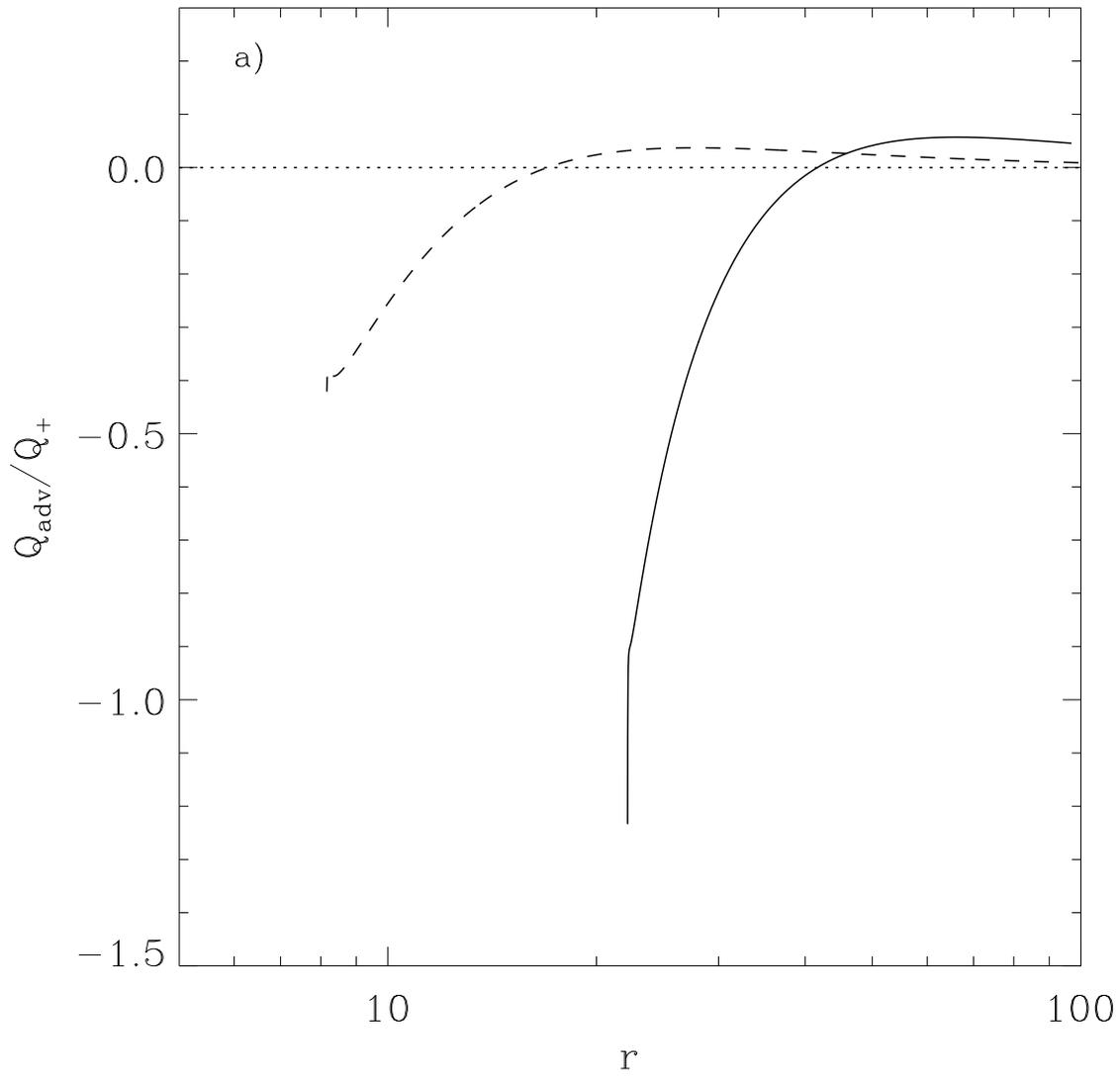
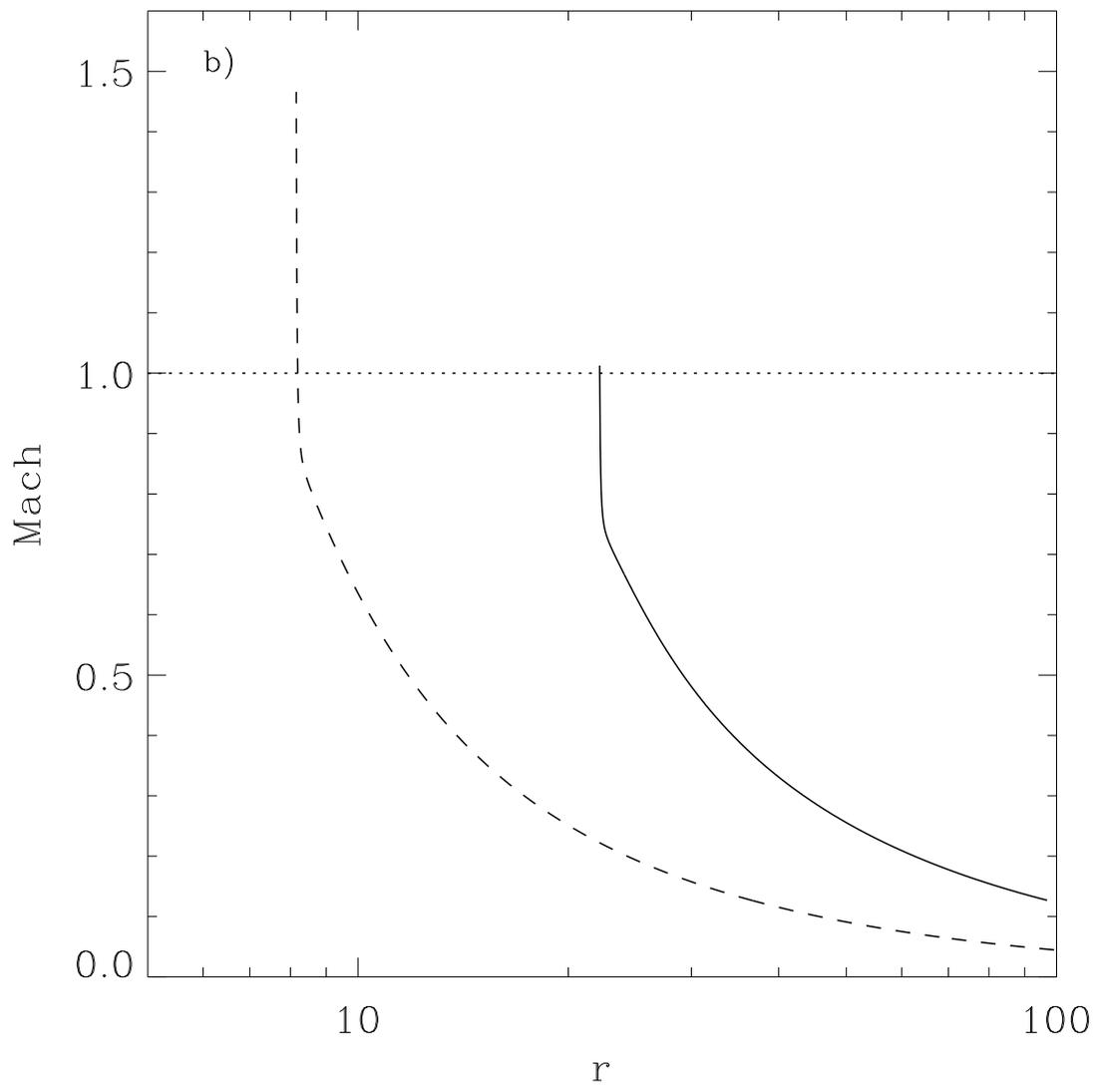
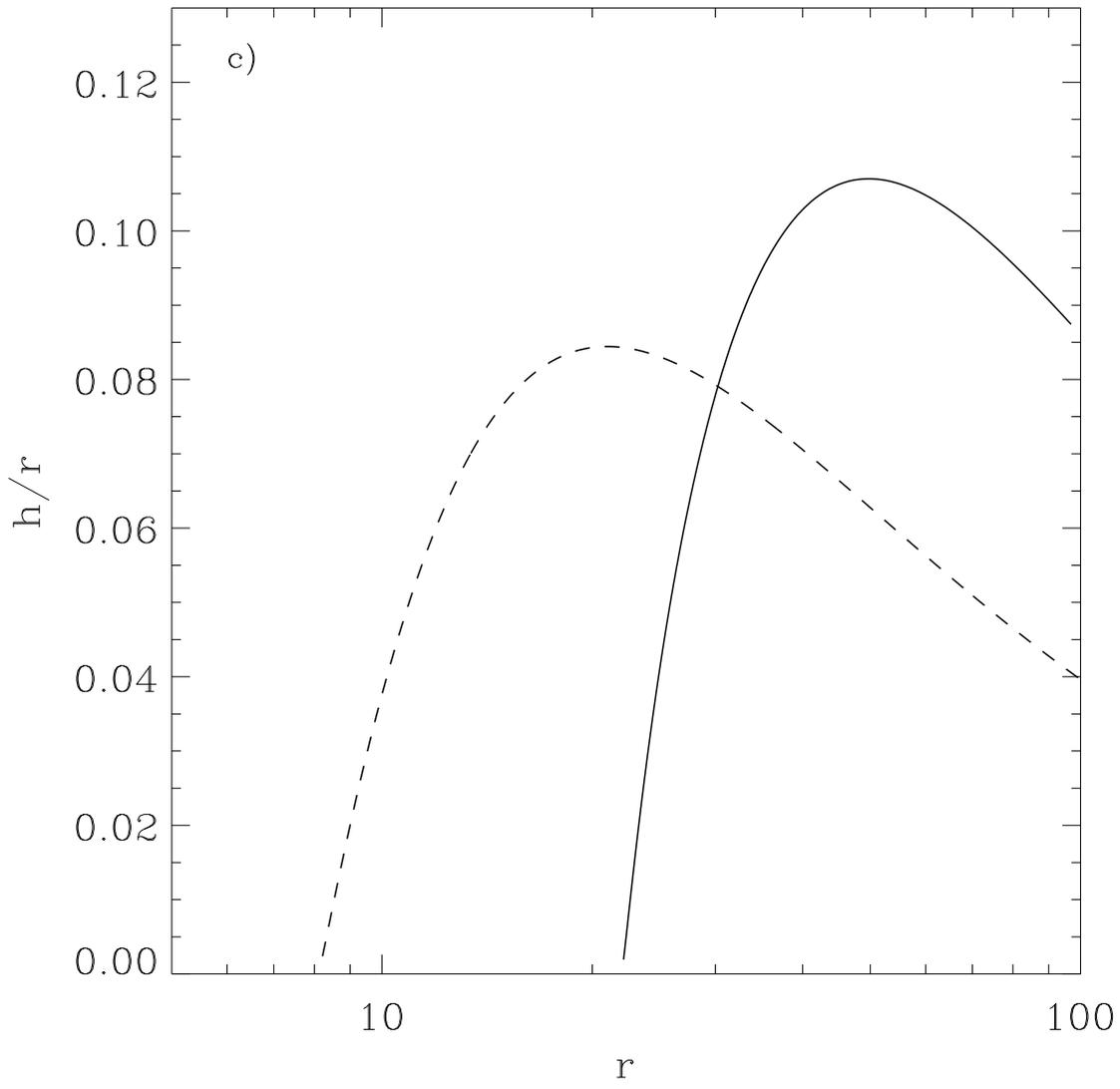
\end{document}